\documentclass[aps,pra,twocolumn]{revtex4}
\usepackage{amsmath}
\usepackage{amssymb}
\usepackage{graphics}

\newcommand\dg{^\dagger}

\newcommand\hc{\mathrm{h.c.}}

\newcommand\half{\frac{1}{2}}
\newcommand\beq{\begin{equation}}
\newcommand\eeq{\end{equation}}
\newcommand\beqa{\begin{eqnarray}}
\newcommand\eeqa{\end{eqnarray}}

\newcommand\qo{\hat q}
\newcommand\po{\hat p}
\newcommand\fo{\hat f}
\newcommand\aop{\hat a}
\newcommand\bo{\hat b}

\newcommand\So{\hat S}
\newcommand\psio{\hat \psi}
\newcommand\phio{\hat \phi}
\newcommand\vphio{\hat \varphi}

\newcommand\Xo{\hat X}
\newcommand\Yo{\hat Y}

\newcommand\Tc{\mathcal{T}}
\newcommand\Nc{\mathcal{N}}
\newcommand\Wc{\mathcal{W}}
\newcommand\Lc{\mathcal{L}}

\newcommand\Fc{\mathcal{F}}
\newcommand\Oc{\mathcal{O}}
\newcommand\Cc{\mathcal{C}}
\newcommand\Dc{\mathcal{D}}
\newcommand\Om{\Omega}

\begin{document}
\title{General Wick's Theorem for bosonic and fermionic operators}   
\author{L. Ferialdi}
\email{ferialdi@ts.infn.it}
\affiliation{Department of Physics, University of Trieste, Strada Costiera 11, 34151 Trieste, Italy\\ Istituto Nazionale di Fisica Nucleare, Trieste Section, Via Valerio 2, 34127 Trieste, Italy}
\author{L. Di\'osi}
\email{diosi.lajos@wigner.hu}
\affiliation{Wigner Research Center for Physics, H-1525 Budapest 114, P.O.Box 49, Hungary\\
Department of Physics of Complex Systems, E\"otv\"os Lor\'and University, Budapest, Hungary}
\begin{abstract}
Wick's theorem provides a connection between time ordered products of bosonic or fermionic fields, and their normal ordered counterparts. We consider a generic pair of operator orderings and we prove, by induction, the theorem that relates them. We name this the General Wick's Theorem (GWT) because it carries Wick's theorem as special instance, when one applies the GWT to time and normal orderings. We establish the GWT both for bosonic and fermionic operators, i.e. operators that satisfy c-number commutation and anticommutation relations respectively. We remarkably show that the GWT is the same, independently from the type of operator involved. By means of a few examples, we show how the GWT helps treating demanding problems by reducing the amount of calculations required.
\end{abstract}
\maketitle

\section{Introduction}
Ever since the advent of quantum mechanics various rules of operator orderings
have been considered, e.g., in canonical quantization \cite{Wey27}, phase space representation \cite{Wig32}, field theory \cite{Dys49,HouKin49,Wic50,And54,Mat55}, quantum optics \cite{CahGla69}, statistical physics~\cite{Kel65}.
Time-ordering $\Tc $ 
and normal ordering $\mathcal{N}$ of quantized fields $\phio(x)$ 
are paradigmatic in relativistic quantum field theory.  
These orderings are related by Wick's theorem \cite{Wic50}, whose conception proved crucial essentially in any area of theoretical physics, because it allows for calculating the matrix elements of the (time-ordered) evolution operator.
Wick's theorem for exponential test functionals 
of a free bosonic/fermionic field $\phio(x)$ can be written into the compact form \cite{Choetal85}:
\beq\label{WT}
\Tc [e^{\int \lambda(x)\phio(x)d^4x}]=e^{C}\mathcal{N}[e^{\int \lambda(x)\phio(x)d^4x}],
\eeq
where $\lambda(x)$ is an arbitrary c-number field and $C$ is its quadratic
functional 
\beq
C=\frac{1}{2}\int\int C(x,y)\lambda(x)\lambda(y)d^4xd^4y\;,
\eeq
where $C(x,y)$ is the kernel of \emph{Wick's contraction}:
\beq\label{ContrTN}
C(x,y)=\Tc \phio(x)\phio(y)-\mathcal{N}\phio(x)\phio(y)\equiv(\Tc -\mathcal{N})\phio(x)\phio(y).
\eeq
Two further forms~\cite{And54,Mat55} of Wick's original theorem were proven
for arbitrary functionals $\Fc$ instead of the exponential test functionals in Eq.\eqref{WT}:
\beqa\label{WTnew1}
\Tc [\Fc(\phio)]&=&\Nc[\Fc(\phio')]\\
\label{WTnew2}\Tc [\Fc(\phio)]&=&e^\Gamma\Nc[\Fc(\phio)] 
\eeqa
where  $\phio'(x)=\phio(x)+\int C(x,y)\frac{\delta}{\delta\phi(y)}d^4y$, and $\Gamma$ is
the quadratic form of the functional derivatives:
\beq
\Gamma=\half\int\int C(x,y)\frac{\delta}{\delta\phi(x)}\frac{\delta}{\delta\phi(y)}d^4xd^4y\,.
\eeq 

Wick's seminal work was extended in different ways, so that the literature of ``generalized Wick's theorem'' includes very different kinds of generalizations, such as: spin chains~\cite{PerCap77,Peretal84}; generalized normal order in quantum chemistry~\cite{KutMuk97,Konetal10}; thermal field theory~\cite{Gau60,Vagetal90,EvaSte96}; phase-space representation of time ordering against a generic ordering~\cite{AgaWol69,AgaWol70}; non-equilibrium Green's functions~\cite{Hal75}; multiphonon theory~\cite{SilPie82}.

We follow the direction first taken  in~\cite{Dio18}, where it was suggested that the theorem in Eqs.~\eqref{WT}-\eqref{ContrTN} still holds if $\Tc$ and $\Nc$ are replaced by any pair  $\Oc$, $\Oc'$ of generic orderings. With these replacements, Eq.~\eqref{WT} is called the \emph{General Wick's Theorem} (GWT), while Eq.~\eqref{ContrTN} defines the general contraction. A tentative proof of GWT for bosonic operators was provided in~\cite{Dio18}, which was circumstancial but gave a strong indication of its correctness.

In this paper, not only we confirm the correctness of the intuition in~\cite{Dio18} for GWT with bosonic operators, but our major
new result is the  unique form of GWT for the pair of generic 
orderings  for bosonic and fermionic operators together. 
Precisely, we consider the forms~\eqref{WTnew1}-\eqref{WTnew2} of Wick's theorem, and we prove the ultimate form of GWT:
\beqa\label{GWT21}
\Oc[\Fc(\phio)]&=&e^\Gamma\Oc'[\Fc(\phio)]\nonumber\\
C(x,y)&=&(\Oc-\Oc')\phio(x)\phio(y)\,,
\eeqa
valid no matter if $\phio$'s are bosonic or fermionic or both together. 

The paper is organized as follows: in Secs. II and III we respectively introduce the definitions of operator orderings and contractions. In Sec. IV we prove the GWT for bosonic operators by induction, while in Sec. V we outline the proof for fermionic operators. In Sec. VI we provide two applications of the GWT, namely to the Baker-Campbell-Hausdorff formula and to quadratic forms, and in Sec. VII we draw our conclusions.

\section{Operator orderings}
We consider a set of operators $\phio_\alpha$ with $\alpha$ belonging to some ordered index set $\Om$.
An ordering operator $\Oc$ rearranges the elements of an input product of operators, to output a suitably ordered one. Orderings $\Oc$ can be bosonic or fermionic, 
defined by 
 \beq\label{Ord}
\Oc[\phio_1\dots\phio_n]=(\pm1)^P \phio_{p_1}\dots\phio_{p_n}\,,
\eeq
when for simplicity's sake the index set $\Om$ consists of integers. Here $P$ is the number of permutations that bring the initial string of indexes $1\dots n$ to the ordered one $p_1\succ p_2\succ\dots\succ p_n$. The signature $\pm1$ distinguishes bosonic orderings ($+1$) from fermionic ones ($-1$). Accordingly, fermionic orderings are sensitive to the order of the operators in the input string, but bosonic orderings are not.
By bosonic/fermionic operators we mean any set of operators satisfying c-number commutation/anticommutation relations: $[\phio_\alpha,\phio_\beta]_\pm\in\mathbb{C}$ 
(where $[\cdot,\cdot]_-=[\cdot,\cdot]$ and $[\cdot,\cdot]_+=\{\cdot,\cdot\}$). The identity~\eqref{Ord} can be trivially extended to composite indices, the paradigmatic example being the time ordering of fields $\phio(x)=\phio(t,\boldsymbol{x})$. We take the occasion to clarify an issue that is often overlooked. 
In our definition \eqref{Ord} we follow Wick who observed that validity of his theorem required the introduction  of the sign $(-1)^P$ for fermionic orderings $\Tc,\Nc$ \cite{Wic50}. So, Wick's time-ordering of fermionic fields differ from Dysons's \cite{Dys49} which does not contain the sign $(-1)^P$.   
Accordingly, fermionic Wick's theorem in general cannot be applied to Dyson-ordered evolution operators for fermionic systems (see e.g.~\cite{Fer17}). In quantum electrodynamics this constitutes no issue because the electromagnetic field couples to the current which is local quadratic in the fields: Dyson's and Wick's orderings coincide in this case.

When we talk about a different ordering $\Oc'$ of the same product $\phio_1\dots\phio_n$, it may be a different permutation of the field operators, but
it will be more general than that. We assume that the operators $\phio_\alpha$
are linear combinations of operators $\vphio_k$, with $k$ possibly belonging to some different index set $\Om'$:
\beq\label{phitov}
\phio_\alpha=\Lc_{\alpha k} \vphio_k\,, 
\eeq
where and henceforth we assume the Einstein convention for repeated indexes, with the additional condition that sums always run on all the elements of the respective index sets.
Here we assume that $\Om$ and $\Om'$  are discrete sets but the following analysis holds true invariably for continuous sets, provided that sums are replaced by intergrals, functions become functionals, matrices become kernels, and partial derivatives are replaced by functional derivatives.
We postulate that $\Oc'$ orders the products of the operators $\vphio_k$.
To be as general as possible, $\Oc$ orders (products of) $\phio_\alpha$'s, cf. Eq.~\eqref{Ord}, but it does not order $\vphio_k$'s. Similarly, $\Oc'$ orders $\vphio_k$'s, but does not order $\phio_\alpha$'s.  A simple example to keep in mind is $\{\phio\}=\qo,\po$, $\Oc=qp$-ordering, $\{\vphio\}=a,a^\dag$, $\Oc'=\mathcal{N}$. Still, we define $\Oc'$-ordering of the $\phio$'s indirectly, using
the expansion \eqref{phitov}:
\beq\label{Ordpr}
\Oc'[\phio_{\alpha_1}\dots\phio_{\alpha_n}]\equiv\left(\prod_{i=1}^n\Lc_{\alpha_ik_i}\right)
\Oc'[\vphio_{k_1}\dots\vphio_{k_n}]\,.
\eeq
The assumption of the linear relationship \eqref{phitov} allows for a simpler and
more transparent proof of GWT compared to the tentative proof in \cite{Dio18}. 
Then the result can shortly be extended for the case of the more generic, implicit linear relationship (cf.~\eqref{linear}).  

\section{Contractions}
Given the pair of  orderings $\Oc,\Oc'$  interpreted in Sec. II, 
we define the matrix of their contraction:
\beq\label{Contr}
\Cc_{\alpha\beta}=(\Oc-\Oc')\phio_\alpha\phio_\beta,
\eeq
 which is symmetric/antisymmetric if the $\phio_\alpha$'s are bosonic/fermionic,
 respectively.
Using Eqs.~\eqref{phitov}-\eqref{Ordpr} we can detail the rhs as follows:
\beqa\label{Ctheta}
\Cc_{\alpha\beta}&=&\theta_{l\succ k}\Lc_{\alpha k}\Lc_{\beta l}[\vphio_k,\vphio_l]_\pm
- \theta_{\beta\succ\alpha}[\phio_{\alpha},\phio_\beta]_\pm\nonumber\\
                                  &=&\left(\theta_{\alpha\succ\beta}-\theta_{k\succ l}\right)[\phio_{\alpha},\phio_\beta]_\pm\,,
\eeqa
where we introduced the step function: $\theta_{\beta\succ\alpha}=1$ if $\beta\succ\alpha$ and zero if $\alpha\succ\beta$. 
About our compact notations in the second line, 
we stress that $\Oc'$-ordering of  the operators $\phio_{\alpha}$ and $\phio_{\beta}$ refers to their
expansion \eqref{phitov} in terms of the operators $\vphio_k$ and $\vphio_l$, respectively. 
When $\Oc=\Tc $ and $\Oc'=\Nc$ to order quantum fields $\phio(x)$ 
the index set $\Om=\{x\}$ becomes continuous, and our generalized contraction \eqref{Contr} yields  Wick's contraction \eqref{ContrTN}
as it should.  Along our forthcoming derivations we need the matrix $
\tilde{\Cc}_{kl}$
of contraction in terms of the $\vphio_k$'s:
\beq\label{Contrvphio}
\tilde{\Cc}_{kl}=(\Oc-\Oc')\vphio_k\vphio_l,
\eeq
satisfying $\Lc_{\alpha k}\Lc_{\beta l} \tilde{\Cc}_{kl}=\Cc_{\alpha\beta}$.

Contractions exist also when the operators $\phio_\alpha$ cannot be 
expressed by linear combinations of the $\vphio_k$'s  like in Eq.~\eqref{phitov}
but the following implicite linear relationship 
with some coefficients $\lambda_\alpha$ and $\tilde{\lambda}_k$
does exist for them (cf. \cite{Dio18}):
\beq\label{linear}
\lambda_\alpha \phio_\alpha=\tilde{\lambda}_k\vphio_k\,(\equiv\Xo)\,
\eeq
where the last identity simply defines the operator $\Xo$.
This is essentially a generalization of~\eqref{phitov}, 
retaining the minimal requirement of linear relationship 
between the $\Oc$-ordered $\phio_\alpha$'s and the $\Oc'$-ordered $\vphio_k$'s.
A simple example to keep in mind is: $\{\phio\}=\{\fo_1+\fo_2,\fo_3\}$, $\{\vphio\}=\{\fo_1,\fo_2+\fo_3\}$; a relation of the type~\eqref{phitov} cannot be established between $\phio$ and $\vphio$, but~\eqref{linear} holds.
The contraction between $\Oc$ and $\Oc'$ can now be established in two steps.
First, we consider the contraction between the trivial ``ordering''  of the lonely 
operator $\Xo$ and the $\Oc$-ordering of the  $\phio_\alpha$'s.
Note that that $\Xo=\lambda_\alpha\phio_\alpha$, which corresponds to Eq.~\eqref{phitov}, 
so the previously defined contraction \eqref{Contr} applies, and it applies
similarly between the trivial ordering of $\Xo$ and the $\Oc'$-ordering of
the $\vphio_k$'s:
\beqa
\Cc^{X\phi}&=&\Xo^2-\Oc\Xo^2\\
\Cc^{X\varphi}&=&\Xo^2-\Oc'\Xo^2.
\eeqa
Second, we obtain the contraction between $\Oc$ and $\Oc'$:
\beq\label{CXX}
\Cc=\Cc^{X\varphi}-\Cc^{X\phi}=(\Oc-\Oc')\Xo^2=
\Cc_{\alpha\beta}\lambda_\alpha\lambda_\beta.
\eeq
In the special case when the explicit relationship \eqref{phitov} holds, the
above equation determines the matrix $\Cc_{\alpha\beta}$ uniquely and
in accordance with \eqref{Contr}, otherwise we shall rely on the scalar
contraction $\Cc$.

\section{Bosonic GWT}
We aim at proving that the $\Oc$-ordering of any function $F$ of operators $\phio_\alpha$, with shorthand notation $F(\phio)$, can be rewritten as the $\Oc'$-ordering of the same function of new operators $\phio'_\alpha$, namely:
\beq\label{GWT1}
\mathcal{O}[F(\phio)]=\mathcal{O}'[F(\phio')]\,,
\eeq
with
\beq\label{Phio}
\phio'_\alpha\equiv\phio_\alpha+\Cc_{\alpha\beta}\partial_\beta\,,
\eeq
$\Cc_{\alpha\beta}$ is the matrix of contraction \eqref{Contr},  and $\partial_\beta=\partial/\partial\phi_\beta$ is a standard c-number derivative
(see Appendix for further details on the specific meaning of such derivatives). 
On the rhs of  GWT \eqref{GWT1} it is to be understood, 
that before  the $\Oc'$-ordering we
express the operators $\phio_\alpha'$ in terms of the operators $\vphio_k'$, cf. \eqref{Ordpr}.  Exploiting \eqref{phitov} we
can write 
\beq\label{phiprtov}
\phio_\alpha'=\Lc_{\alpha k}\vphio_k'
\eeq
\beq\label{vphipr}
\vphio_k'=\vphio_k+\tilde{\Cc}_{kl}\tilde{\partial}_l
\eeq
with notation $\tilde{\partial}_l=\partial/\partial\varphi_{l}$, and where we recall that the matrix  $\tilde{\Cc}_{kl}$ satisfies
$\Lc_{\alpha k}\Lc_{\beta l} \tilde{\Cc}_{kl}=\Cc_{\alpha\beta}$.
Since any operator functional can be expanded in power series, we will work with products of operators. 
We prove the GWT by induction: we assume that
\beq\label{nGWT}
\Oc\left[\prod_{i=1}^n\phio_{\alpha_i}\right]=\Oc'\left[\prod_{i=1}^n\phio'_{\alpha_i}\right]
\eeq
holds up to a given $n$ (the cases $n=0$ and $n=1$ are trivially true), and we prove that
\beq\label{n+1GWT}
\Oc\left[\phio_{\alpha}\prod_{i=1}^n\phio_{\alpha_i}\right]=\Oc'\left[\phio'_{\alpha}\prod_{i=1}^n\phio'_{\alpha_i}\right]\,.
\eeq
Let us assume that $\Oc$ orders the operators $\phio$ with decreasing index from left to right, i.e. $\alpha_n\succ\dots\succ\alpha_1$:
\beq\label{On}
\Oc\left[\prod_{i=1}^n\phio_{\alpha_i}\right]=\phio_{\alpha_n}\dots\phio_{\alpha_1}\,,
\eeq
We can thus rewrite the lhs of Eq.~\eqref{n+1GWT} as follows:
\beq\label{On+1}
\Oc\left[\phio_{\alpha}\prod_{i=1}^n\phio_{\alpha_i}\right]=\phio_{\alpha_n}\dots\phio_{\alpha_{j+1}}\phio_{\alpha}\phio_{\alpha_j}\dots\phio_{\alpha_1}\,,
\eeq
where $\alpha_n\succ\dots\succ\alpha_{j+1}\succ\alpha\succ\alpha_j\succ\dots\succ\alpha_1$.
In order to be able to exploit Eq.~\eqref{nGWT}, we need to bring $\phio_{\alpha}$ outside the product, and we arbitrarily choose to do so by moving $\phio_{\alpha}$ to the left (needless to say, the GWT can be equivalently proved also by moving $\phio_{\alpha}$ to the right). This can be done by exploiting the following identity
\beqa\label{switch}
\phio_{\alpha_n}\dots\phio_{\alpha_{j+1}}\phio_{\alpha}\phio_{\alpha_j}\dots\phio_{\alpha_1}
&=&\phio_{\alpha_n}\dots\phio_{\alpha}\phio_{\alpha_{j+1}}\phio_{\alpha_{j}}\dots\phio_{\alpha_1}\nonumber\\
&&\hspace{-1.5cm}-[\phio_{\alpha},\phio_{\alpha_{j+1}}]\partial_{\alpha_j+1}\Oc\left[\prod_{i=1}^n\phio_{\alpha_i}\right]\,,
\eeqa
where $\alpha_{j+1}$ is the label of $\alpha$'s left neighbour, Einstein summation does not apply to it.
When iterated, this identity leads to\beqa\label{left}
\Oc\left[\phio_{\alpha}\prod_{i=1}^n\phio_{\alpha_i}\right]&=&\phio_{\alpha}\,\mathcal{O}\left[\prod_{i=1}^n\phio_{\alpha_i}\right]\nonumber\\
&&- \theta_{\beta\succ\alpha}[\phio_{\alpha},\phio_\beta]\partial_\beta \,\mathcal{O}\left[\prod_{i=1}^n\phio_{\alpha_i}\right]\,,
\eeqa
where the Einstein convention is reactivated. 

We apply Eq.~\eqref{nGWT}, i.e. GWT up to order $n$, yielding
\beqa\label{nGWTused}
\mathcal{O}\left[\phio_{\alpha}\prod_{i=1}^n\phio_{\alpha_i}\right]&=&\phio_{\alpha}'\,\mathcal{O}'\left[\prod_{i=1}^n\phio_{\alpha_i}'\right]
                                                                                                                                             -\Cc_{\alpha\beta}\partial_\beta\Oc'\left[\prod_{i=1}^n\phio_{\alpha_i}'\right]\nonumber\\
&&- \theta_{\beta\succ\alpha}[\phio_{\alpha},\phio_\beta]\partial_\beta \,\mathcal{O}'\left[\prod_{i=1}^n\phio_{\alpha_i}'\right]\,,
\eeqa
where we inserted $\phio_{\alpha}=\phio_{\alpha}'-C_{\alpha\beta}\partial_\beta$.
Let us concentrate on the first term on the rhs, onto which we exploit Eq.~\eqref{phiprtov} in order to prepare for $\Oc'$-ordering :
\beq\label{firstterm}
\phio_{\alpha}'\,\Oc'\left[\prod_{i=1}^n\phio_{\alpha_i}'\right]
=\Lc_{\alpha k}\left[\prod_{i=1}^n\Lc_{\alpha_ik_i}\right]
\vphio_k'\Oc' \left[\prod_{i=1}^n\vphio_{k_i}'\right]\,.                                                                                                                                          
\eeq
The operator part on the rhs can further be written as
\beqa
\vphio_k'\Oc' \left[\prod_{i=1}^n\vphio_{k_i}'\right]&=&\Oc' \left[ \vphio_k'\prod_{i=1}^n\vphio_{k_i}'\right]\nonumber\\
&&+\theta_{l\succ k}[\vphio_k',\vphio_l']\tilde{\partial}_l\Oc' \left[\prod_{i=1}^n\vphio_{k_i}'\right]\,,
\eeqa
so that by re-using Eq.~\eqref{phiprtov}, Eq.~\eqref{firstterm} becomes
\beqa
\phio_{\alpha}'\,\Oc'\left[\prod_{i=1}^n\phio_{\alpha_i}'\right]
&=&\Oc' \left[ \phio_\alpha'\prod_{i=1}^n\phio_{\alpha_i}'\right]\nonumber\\
&&\hspace{-1cm}+\Lc_{\alpha k}\theta_{l\succ k}[\vphio_k',\vphio_l']\tilde{\partial}_l\Oc' \left[\prod_{i=1}^n\phio_{\alpha_i}'\right]
\eeqa
With this, we can write Eq.~\eqref{nGWTused} into this form:
\beq
\Oc\left[\phio_{\alpha}\prod_{i=1}^n\phio_{\alpha_i}\right]=\Oc' \left[ \phio_\alpha'\prod_{i=1}^n\phio_{\alpha_i}'\right]
+\Dc \Oc' \left[\prod_{i=1}^n\phio_{\alpha_i}'\right]
\eeq
where $\Dc $ is the following differential operator:
\beq
\Dc =\Lc_{\alpha k}\theta_{l\succ k}[\vphio_k',\vphio_l']\tilde{\partial}_l-\Cc_{\alpha\beta}\partial_\beta
- \theta_{\beta\succ\alpha}[\phio_{\alpha},\phio_\beta]\partial_\beta
\eeq
At this stage we substitute the identity $\tilde{\partial}_l=\Lc_{\beta l}\partial_\beta$ and,
invoking the definiton \eqref{vphipr}, we can replace $[\vphio_k',\vphio_l']=[\vphio_k,\vphio_l]+\tilde{\Cc}_{kl}-\tilde{\Cc}_{lk}=[\vphio_k,\vphio_l]$, yielding
\beqa
\Dc &=&\left[\Lc_{\alpha k}\Lc_{\beta l}\theta_{l\succ k}[\vphio_k,\vphio_l]
- \theta_{\beta\succ\alpha}[\phio_{\alpha},\phio_\beta]
-\Cc_{\alpha\beta}
\right]\partial_\beta\nonumber\\
                       &=&\left[(\theta_{\alpha\succ\beta}-\theta_{k\succ l})[\phio_{\alpha},\phio_\beta]
-\Cc_{\alpha\beta}
\right]\partial_\beta\,.
\eeqa
Induction is done and the GWT \eqref{GWT1} is proven provided $\Dc$ vanishes, and this happens if we use the form \eqref{Ctheta} with commutator of the
contraction $\Cc_{\alpha\beta}$. The GWT~\eqref{GWT1}-\eqref{Phio} is thus proven.

Now we turn toward the proof of the ultimate form \eqref{GWT21} of the GWT.
There is a transformation of equivalence between each $\phio'_\alpha$ and
$\phio_\alpha$:
\beq\label{Aprime}
\phio'_{\alpha}=e^{\Gamma}\phio_{\alpha} e^{-\Gamma}\,,
\eeq
where
\beq\label{Gamma}
\Gamma= \frac{1}{2}\Cc_{\alpha\beta}\partial_\alpha\partial_\beta\,.
\eeq
Equation \eqref{Aprime} can be confirmed, e.g., by Taylor expanding
the rhs:
\beq
\phio_{\alpha}+[\Gamma,\phio_{\alpha}]+\frac{1}{2}[\Gamma,[\Gamma,\phio_{\alpha}]]+\dots\,,
\eeq
and observing that $[\Gamma,\phio_{\alpha}]=\Cc_{\alpha\beta}\partial_\beta$, while higher order commutators are zero.
This equivalence transformation allows us to write the GWT~\eqref{GWT1} in the following way:
\beq\label{GWT2}
\Oc[F(\phio)]=e^{\Gamma}\Oc'[F(\phio)]\,,
\eeq
where $e^{-\Gamma}$ was dropped because the derivatives have nothing to act upon.

The above proofs of the two forms \eqref{GWT1} and \eqref{GWT2} of bosonic GWT
required the explicit linear relationship \eqref{phitov}. Here we are going to show
that a similar GWT exists if \eqref{phitov} does not hold, but the 
weaker, implicite linear relationship \eqref{linear} does. For this case we
defined the contraction $\Cc=(\Oc-\Oc')\Xo^2$ by Eq.~\eqref{CXX}. The GWT expresses $\Oc$-ordering of a function $F$ in terms of $\Oc'$-ordering
of $F$ therefore  it must be possible to express $F$ both in terms of $\phio$ and of $\vphio$. Accordingly, here we need to restrict the functions $F$ to the class $F(\lambda_\alpha\phio_\alpha)=F(\Xo)$, 
that is an unavoidable compromise when~\eqref{phitov} does not hold.

We are going to derive the form \eqref{GWT2} of GWT
between orderings $\Oc$ and $\Oc'$ in two steps, according to those in Sec. III.
First, since $\Xo$ is a linear combination of the $\phio_\alpha$'s, as well as of
the $\vphio_k$'s, we can apply GWT \eqref{GWT2} between the trivial ``ordering''  of the lonely operator $\Xo$ and the $\Oc$-ordering of the  $\phio_\alpha$'s,
and between $\Xo$ and the $\vphio_k$'s as well:
\beqa
F(\Xo)&=&e^{\frac{1}{2}\Cc^{X\phi}\partial_X^2}\Oc [F(\Xo)]\nonumber\\
F(\Xo)&=&e^{\frac{1}{2}\Cc^{X\varphi}\partial_X^2}\Oc' [F(\Xo)]
\eeqa
Second, we get from here the GWT, extended from the case \eqref{phitov} to  \eqref{linear}: 
\beq\label{GWT2X}
\Oc [F(\Xo)]=e^{\frac{1}{2}\Cc\partial_X^2}\Oc' [F(\Xo)]\,.
\eeq
Note that, invoking the second expression of $\Cc$ in \eqref{CXX} and
using the chain rule of derivatives, we rewrite Eq.(\ref{GWT2X}) as follows:
\beq\label{GWT3X}
\Oc [F(\Xo)]=e^{\frac{1}{2}\Cc_{\alpha\beta}\partial_\alpha\partial_\beta}\Oc' [F(\Xo)]
\eeq
which is the same form \eqref{Gamma}-\eqref{GWT2} that we derived
for the direct relationship \eqref{phitov} except for the mentioned restriction
on the form of the function $F$. We remark that if relation~\eqref{linear} holds for different sets (labelled by superscript $i$) $\{\lambda_\alpha^i\}$, then the GWT~\eqref{GWT3X} holds for $F(\Xo^1,\Xo^2,\dots)$ with $\Xo^i=\lambda_\alpha^i\phio_\alpha$. Eventually, if Eq.~\eqref{linear} holds for any choice of $\{\lambda_\alpha^i\}$, then~\eqref{phitov} becomes existing and~\eqref{GWT3X} yields Eq.~\eqref{GWT2}, valid for $F(\phio)$ without restriction.

We add, for completness, that on exponential test functions $F(\Xo)=e^{\Xo}$, our Eq.~\eqref{GWT3X} yields
\beq\label{GWTX}
\Oc [e^{\Xo}]=e^{\frac{1}{2}\Cc_{\alpha\beta}\lambda_\alpha\lambda_\beta}\Oc' [e^{\Xo}]
\eeq
which is the form of GWT proposed in~\cite{Dio18} to generalize the form 
\eqref{WT}-\eqref{ContrTN} of Wick's theorem~\footnote{We take the opportunity to correct two typos in~\cite{Dio18}. Expression~(6) of contraction $C$ should contain a factor 1/2 in front of the integral. The contraction (42) should read $C_t=(i\hbar/m)\int_0^t d\tau\int_0^{\tau}ds \,\tau F_{\tau} F_s$.}.

\section{Fermionic GWT}
The advantage of the functional approach is that it allows to extend the GWT to fermionic systems. Namely, we aim at proving the GWT~\eqref{GWT1}-\eqref{Phio}, where now $\phio$'s (and $\vphio$'s) are fermionic operators (in the sense defined in Sec.~II). We therefore retain the same setting as Secs.~II-IV, in particular Eqs.~\eqref{phitov},~\eqref{Ctheta},~\eqref{nGWT} and~\eqref{On}, and we prove the GWT by induction. We recall that fermionic orderings depend on the initial order of the operators. In what follows, we nonetheless retain the simple notation $\prod_i \phio_i$ for products of
fermionic operators, which now denotes a definite initial ordering whose choice
is arbitrary, and anyway cancels from the GWT. 
The equation corresponding to Eq.~\eqref{On+1} for fermions reads:
\beq\label{Ofer}
\mathcal{O}\left[\phio_{\alpha}\prod_{i=1}^n\phio_{\alpha_i}\right]=(-1)^{n-j}\phio_{\alpha_n}\dots\phio_{\alpha_{j+1}}\phio_{\alpha}\phio_{\alpha_j}\dots\phio_{\alpha_1}\,,
\eeq
where the factor $(-1)^{n-j}$ is due to the fact that the fermionic ordering brings a factor $(-1)$ for each permutation. In order to move $\phio_\alpha$ to the left we need to rewrite the rhs of this equation by iterating the following identity
\beqa\label{idfer}
\phio_{\alpha_n}\dots\phio_{\alpha_{j+1}}\phio_{\alpha}\phio_{\alpha_j}\dots\phio_{\alpha_1}
\!\!&=&\!\!-\phio_{\alpha_n}\dots\phio_{\alpha}\phio_{\alpha_{j+1}}\phio_{\alpha_{j}}\dots\phio_{\alpha_1}\nonumber\\
&&\hspace{-2.5cm}-(-1)^{n-j}\{\phio_{\alpha},\phio_{\alpha_{j+1}}\}\partial_{\alpha_j+1}\Oc\left[\prod_{i=1}^n\phio_{\alpha_i}\right]\,,
\eeqa
where the partial derivative is  the standard Grassmann derivative, detailed in the Appendix.
We can thus rewrite Eq.~\eqref{Ofer} as follows
\beqa\label{leftfer}
\mathcal{O}\left[\phio_{\alpha}\prod_{i=1}^n\phio_{\alpha_i}\right]&=&\phio_{\alpha}\,\mathcal{O}\left[\prod_{i=1}^n\phio_{\alpha_i}\right]\\
&&- \theta_{\beta\succ\alpha}\{\phio_{\alpha},\phio_\beta\}\partial_\beta \,\mathcal{O}\left[\prod_{i=1}^n\phio_{\alpha_i}\right]\nonumber\,.
\eeqa
which confirms that the initial ordering of operators is not influent for the proof of GWT. Remarkably, this equations has exactly the same structure as Eq.~\eqref{left} for bosons. Accordingly, from here the proof of the fermionic GWT follows the lines of the bosonic one 
and we will not repeat it here. The final result is given by Eqs.~\eqref{GWT1}-\eqref{Phio} with the anticommutator form~\eqref{Ctheta} for the contraction. We further remark that the form~\eqref{GWT2} of GWT holds also for fermions, with the same definition~\eqref{Gamma} for $\Gamma$.

An explicit form of GWT for fermionic fields $\psio$ and $\psio^\dag$ can be easily obtained by considering the set $\{\phio_\alpha\}_{\alpha=1}^{2n}=\{\psio_1,\psio_1^\dag,\dots\psio_n,\psio_n^\dag\}$. Since $\{\psio_{\alpha},\psio_\beta\}=\{\psio^\dag_{\alpha},\psio^\dag_\beta\}=0$ and $\{\psio_{\alpha},\psio^\dag_\beta\}\in\mathbb{C}$, by defining
\beq
\bar{\Cc}_{\alpha\beta}=\Big(\theta_{\alpha\succ\beta}-\theta_{k\succ l}\Big)\{\psio_{\alpha},\psio^\dag_\beta\}\,,
\eeq
we observe that $\Cc_{\alpha\beta}=0$ when $\alpha$ and $\beta$ are both even or odd; $\Cc_{\alpha\beta}=\bar{\Cc}_{\alpha\beta}$ when $\alpha$ is odd and $\beta$ is even; $\Cc_{\alpha\beta}=-\bar{\Cc}_{\alpha\beta}$ when $\alpha$ is even and $\beta$ is odd.
Therefore, we can express the GWT for fermionic fields as follows:
\beq\label{ferGWT}
\mathcal{O}[F(\psio,\psio^\dag)]=\mathcal{O}'[F(\psio',\psio'^{\,\dag})]\,,
\eeq
with
\beqa
\psio'_{\alpha}&=&\psio_{\alpha}-\bar{\Cc}_{\alpha\beta}\partial_\beta^\dag\\
\psio'^{\,\dag}_{\alpha}&=&\psio^\dag_{\alpha}+\bar{\Cc}_{\alpha\beta}\partial_\beta\,.
\eeqa
If we identify $\Oc=\mathcal{T}$ and $\Oc'=\mathcal{N}$, we recover the form of fermionic Wick's theorem discussed in~\cite{And54,Mat55}.

\section{Examples}
In this Section we provide some applications of the GWT in order to show how this helps to tackle in a straightforward manner problems that possibly involve long calculations.
Let us start by considering two generic bosonic operators $\Xo$, $\Yo$, and the ordering $\Oc_{XY}$ that pushes the operator $\Xo$ to the left and the operator $\Yo$ to the right, i.e.
\beq
\Oc_{XY}[e^{\Xo+\Yo}]=e^{\Xo}e^{\Yo}
\eeq
We apply the GWT between $\Oc_{XY}$ and the Weyl ordering $\mathcal{W}$ defined by
\beq\label{Weyl}
\mathcal{W}[e^{\Xo+\Yo}]=e^{\Xo+\Yo}.
\eeq
It is strightforward to show that the contraction~\eqref{Contr} reads $\Cc_{XY}=(\Oc_{XY}-\mathcal{W})\Xo\Yo=\half[\Xo,\Yo]$, and the GWT~\eqref{GWT2} predicts
\beq
e^{\Xo}e^{\Yo}=e^{\Xo+\Yo+\half[\Xo,\Yo]}\,.
\eeq
We thus see that the Baker-Campbell-Hausdorff (BCH)~\cite{BCH1,BCH2,BCH3} formula for bosonic operators is a special instance of the GWT. This reverses the point of view taken in~\cite{Dio18}, where the tentative proof of GWT was based on the BCH formula, and therefore the GWT was understood to be a consequence of BCH, not its generalization.

Another interesting example is the application of the GWT to quadratic forms of the type $e^{\half D_{\alpha\beta}\phio_\alpha\phio_\beta}$, which occur e.g. in open quantum systems~\cite{DioFer14,Fer16} and in quantum optics. For the special class where $D$ is real positive (or negative), we are going to show that
\beq\label{GWTquad}
\Oc[e^{\half D_{\alpha\beta}\phio_\alpha\phio_\beta}]
=\sqrt{|D'|/|D|}\Oc'[e^{\half D'_{\alpha\beta}\phio_\alpha\phio_\beta}]\,,
\eeq
with $D'_{\alpha\beta}=(D_{\alpha\beta}^{-1}-\Cc_{\alpha\beta})^{-1}$, 
valid if  $D'>0$ (or negative when $D<0$). 
We introduce the random real Gaussian variables $\xi_\alpha$ of zero mean
$\boldsymbol{M}\xi_\alpha=0$ and correlation 
$\boldsymbol{M}\xi_\alpha\xi_\beta=D_{\alpha\beta}$. The symbol $\boldsymbol{M}$ stands for the Gaussian integral
\beq\label{gaussint}
\boldsymbol{M} f(\xi_\alpha)\equiv\frac{1}{\sqrt{|D|}}\int f(\xi_\alpha)e^{-\half D^{-1}_{\alpha\beta}\xi_\alpha\xi_\beta}
\prod_{\alpha\in\Omega} \frac{d\xi_\alpha}{\sqrt{2\pi}}\,,
\eeq
which allows to write
\beq\label{expmean}
\boldsymbol{M}e^{\xi_\alpha\phio_\alpha}=e^{\half D_{\alpha\beta}\phio_\alpha\phio_\beta}\,.
\eeq
 Then, using the GWT~\eqref{GWT2}, we write
\beqa
\hspace{-0.5cm}e^{\half C_{\alpha\beta}\partial_\alpha\partial_\beta}
\Oc'[e^{\half D_{\alpha\beta}\phio_\alpha\phio_\beta}]
\!\!&=&\!\!e^{\half C_{\alpha\beta}\partial_\alpha\partial_\beta}
\boldsymbol{M}\Oc'[e^{\xi_\alpha\phio_\alpha}]\nonumber\\
&=&\!\!\boldsymbol{M}
\Oc'[e^{\half C_{\alpha\beta}\xi_\alpha\xi_\beta+\xi_\alpha\phio_\alpha}]\,,
\eeqa
and performing the Gaussian integral according to~\eqref{gaussint} we eventually obtain Eq.~\eqref{GWTquad}. The proof for negative $D$ is readily obtained by replacing $\xi_\alpha$ by $i \xi_\alpha$. In absence of the GWT, such a re-ordering of a quadratic form would require applying repeatedly the BCH formula on the lhs of Eq.~\eqref{expmean}, and re-summation of the contributions obtained. It is thus clear that the GWT reduces the amount of calculations required.

The application of the GWT to the single-mode squeezing operator $e^{i\kappa\qo\po}$, which is a special case of the previous example with $D$ indefinite, was considered earlier in \cite{Dio18,Dio18a}.
Here we calculate the $\Nc$-ordered form of the two-mode squeezing operator $\So(g)=\exp(g\aop\bo-\hc)$, where the emission operators of the two modes are
$\aop,\bo$, respectively, and the squeezing parameter $g$ can be chosen non-negative. 
The main steps are similar as before, just we need two independent complex random Gaussians $\xi_1,\xi_2$, with correlations $\boldsymbol{M}\vert\xi_1\vert^2=\boldsymbol{M}\vert\xi_2\vert^2=g$ and
$\boldsymbol{M}\xi_1^2=\boldsymbol{M}\xi_2^2=0$. Then
\beqa
\So(g)&=&\boldsymbol{M}e^{(\xi_1\aop+\xi_1^\star\bo+\xi_2\aop\dg-\xi^\star_2\bo\dg)}\nonumber\\
  \label{Sweyl}            &=&\boldsymbol{M}\Wc [e^{(\xi_1\aop+\xi_1^\star\bo+\xi_2\aop\dg-\xi^\star_2\bo\dg)}],
\eeqa 
where the second identity is simply given by the definition~\eqref{Weyl} of Weyl ordering. We can now apply the GWT~\eqref{GWT2} to write
\beqa
\Wc [e^{(\xi_1\aop+\xi_1^\star\bo+\xi_2\aop\dg-\xi^\star_2\bo\dg)}]=&&\nonumber\\
&&\hspace{-2.5cm}\Nc [e^{\half(\xi_1\xi_2-\xi^\star_1\xi^\star_2)}e^{(\xi_1\aop+\xi_1^\star\bo+\xi_2\aop\dg-\xi^\star_2\bo\dg)}]\,,
\eeqa
where we exploited the fact that the only non null contractions are $(\Wc-\Nc)\aop\aop^\dag=(\Wc-\Nc)\bo\bo^\dag=\half$. By replacing this identity into Eq.~\eqref{Sweyl} and performing the integration we find the squeezing operator in normal ordering:
\beq
\So(g)=\sqrt{(g^2+1)}\Nc [e^{\frac{g}{g^2+1}[\aop\bo-\bo^\dag\aop^\dag-2g(\aop^\dag\aop+\bo^\dag\bo+1)]}]
\eeq
 We remark that the same result might have been obtained directly from Eq.~\eqref{GWTquad} by performing the suitable replacements.
 
 We eventually mention that in a series of papers~\cite{AgaWol70,AgaWol70b,AgaWol70c}, Agarwal and Wolf set up a phase-space method to investigate, among other issues, expectation values of functions of creation and annihilation operators, ordered according to $\Wc$, $\Nc$, and $\mathcal{A}$ (anti-normal ordering). The GWT allows to recover and generalize this result, without the need to resort to the phase-space formalism. Namely, the GWT~\eqref{GWT3X} allows to write
 \beq
 \langle \Oc[F(\{\aop_\gamma\},\{\aop^\dag_\gamma\})]\rangle= \langle e^{\half \Cc_{\alpha\beta}\partial_\alpha\partial^\dag_\beta} \Oc'[F(\{\aop_\gamma\},\{\aop^\dag_\gamma\})]\rangle
 \eeq
where the contraction is $\Cc_{\alpha\beta}=(\Oc-\Oc')\aop_\alpha\aop^\dag_\beta$, and the derivatives are defined as $\partial_\alpha=\partial/\partial a_\alpha$ and $\partial^\dag_\beta=\partial/\partial a^\dag_\beta$. By recognizing that 
\beq
(\Wc-\Nc)\aop_\alpha\aop^\dag_\beta=(\mathcal{A}-\Wc)\aop_\alpha\aop^\dag_\beta=\half(\mathcal{A}-\Nc)\aop_\alpha\aop^\dag_\beta=\frac{\delta_{\alpha\beta}}{2}\,,
\eeq 
one recovers the results in~\cite{AgaWol70c}.

\section{Conclusions}
We have proven the General Wick's Theorem, namely the generalization of Wick's theorem (which relates time ordering to normal ordering) to any pair of operator orderings. We have shown that the GWT has the same form both for bosonic and fermionic operators, i.e. those operators satisfying c-number (anti)commutation relations. As an application, we have demonstrated that the BCH formula for bosonic operators is a special instance of the GWT. We further considered the ordering of quadratic forms and we have shown that the GWT allows to treat it in a rather straightforward manner, sensibly reducing the amount of calculations required with respect to earlier approaches.
The relationship provided by the GWT is so general that it may possibly be applied in any field where operators ordering plays a role.
Whether some form of GWT can be proven for operators satisfying general commutation relations is still an open issue, and it will be subject of future research.

\section*{Acknowledgements}
LF acknowledges financial support from the H2020 FET Project TEQ (grant n. 766900).
LD thanks NRDI ``Frontline'' Research Excellence Programme
(grant no. KKP133827) and research grant (grant. no. K12435).

\section*{Appendix}
In the main text we make extensive use of derivative symbols to which special meanings are attached~\cite{And54,Mat55}, this Appendix aims at clarifying  their meaning and properties. The bosonic derivatives $\partial=\partial/\partial\phi$ used in Sec.~IV should be understood as standard c-number derivatives. 
The action of $\partial/\partial\phi$ on a function of operators is thus equivalent to temporarily treating $\phio$ like a c-number, 
then performing the derivative $\partial/\partial\phi$, and eventually restoring the hat on $\phio$:
\beq
\frac{\partial}{\partial\phi}F(\phio,\psio,\dots,\hat{\zeta})= \left.\frac{\partial}{\partial\phi}F(\phi,\psio,\dots,\hat{\zeta})\right\rvert_{\phi=\phio}\,.
\eeq
As such, bosonic derivatives satisfy the following commutation relations: 
\beq
\left[\partial_\alpha,\phio_\beta\right]=\delta_{\alpha\beta}\,,\qquad\left[\partial_\alpha,\partial_\beta\right]=0\,.
\eeq

Similarly, in Sec.~V, where we are dealing with fermionic operators $\phio$, the derivatives $\partial/\partial\phi$ should be understood as Grassmann derivatives, which satisfy
\beq
\left\{\partial_\alpha,\phio_\beta\right\}=\delta_{\alpha\beta}\,,\qquad\left\{\partial_\alpha,\partial_\beta\right\}=0\,.
\eeq
In order to perform the Grassmann derivative of a product of fermionic operators one first needs to move the operator to be differentiated close to the derivative, by exploiting anticommutation relations, and then the derivative can act as a standard c-number derivative:
\beqa
\partial_{\alpha_j+1}\prod_{i}\phio_{\alpha_i}&=&(-1)^{n-j-1}\partial_{\alpha_j+1}\left(\phio_{\alpha_{j+1}}\prod_{i\neq j+1}\phio_{\alpha_i}\right)\nonumber\\
&=&(-1)^{n-j-1}\prod_{i\neq j+1}\phio_{\alpha_i}\,.
\eeqa
This equation explains the term $(-1)^{n-j}$ in the second line of Eq.~\eqref{idfer}.


\begin{thebibliography}{99}
\bibitem{Wey27} H. Weyl, Z. Phys. {\bf 46}, 1 (1927).
\bibitem{Wig32} E.P. Wigner, Phys. Rev. {\bf 40}, 749 (1932).
\bibitem{HouKin49} A. Hourlet and A. Kind, Helv. Phys. Acta {\bf 22}, 319 (1949).
\bibitem{Dys49} F.J. Dyson, Phys. Rev. {\bf 75}, 486 (1949).
\bibitem{Wic50} G.C. Wick, Phys. Rev. {\bf 80}, 268 (1950).
\bibitem{And54} J. L. Anderson, Phys. Rev. {\bf94}, 703 (1954).
\bibitem{Mat55} T. Matsubara, Prog. Theor. Phys. {\bf 14}, 351 (1955).
\bibitem{CahGla69} K.E. Cahill and R.J. Glauber, Phys. Rev. {\bf 177}, 1857 (1969).
\bibitem{Kel65} L. V. Keldysh, Sov. Phys. JEPT {\bf20}, 1018 (1965).
\bibitem{Choetal85} K. Chou, Zh. Su, B. Hao, and L. Yu, Phys. Rep. {\bf 118} 1 (1985).
\bibitem{PerCap77} J.H.H. Perk, H.W. Capel, Physica A {\bf89}, 265 (1977).
\bibitem{Peretal84} J.H.H. Perk, H.W. Capel, G.R.W. Quispel, and F.W. Nijhoff, Physica A {\bf123}, 1 (1984).
\bibitem{KutMuk97} W. Kutzelnigg, D. Mukherjee, J. Chem. Phys. {\bf 107}, 432 (1997).
\bibitem{Konetal10} L. Kong, M. Nooijen, D. Mukherjee, J. Chem. Phys. {\bf 132}, 234107 (2010).
\bibitem{Gau60} M. Gaudin, Nucl. Phys. {\bf 15}, 89 (1990).
\bibitem{Vagetal90} A. Vaglica , C. Leonardi, G. Vetri, J. Mod. Opt, {\bf 37}, 1487 (1990).
\bibitem{EvaSte96} T.S. Evans, D.A. Steer, Nucl. Phys. B {\bf 474}, 481 (1996).
\bibitem{AgaWol69} G. S. Agarwal, E. Wolf, Lettere Nuovo Cimento,{\bf 1}, 140 (1969). 
\bibitem{AgaWol70} G. S. Agarwal, E. Wolf, Phys. Rev. D, {\bf 2}, 2206 (1970). 
\bibitem{Hal75} A. G. Hall, J. Phys. A: Math. Gen. {\bf8}, 214 (1975).
\bibitem{SilPie82} B. Silvestre-Brac, R. Piepenbring, Phys. Rev. C {\bf26}, 2640 (1982).
\bibitem{Dio18}  L. Di\'osi, J. Phys. {\bf A51}, 365201 (2018).
\bibitem{Fer17} L. Ferialdi Phys. Rev. A {\bf 95}, 020101(R) (2017); {\it ibid.} {\bf 95}, 069908(E) (2017).
\bibitem{BCH1} J. Campbell, Proc. Lond. Math. Soc. {\bf 28}, 381 (1897).
\bibitem{BCH2} H. Baker, Proc. Lond. Math. Soc. {\bf 34}, 347 (1902).
\bibitem{BCH3} F. Hausdorff, Ber. Verh. Saechs. Akad. Wiss. Leipzig {\bf 58}, 19 (1906).
\bibitem{DioFer14} L. Di\'osi, L. Ferialdi, Phys. Rev. Lett. {\bf 113}, 200403 (2014).
\bibitem{Fer16} L. Ferialdi, Phys. Rev. Lett. {\bf 116}, 120402 (2016).
\bibitem{Dio18a} L. Diosi, J. Russ. Las. Res. {\bf39}, 349 (2018).
\bibitem{AgaWol70b} G. S. Agarwal, E. Wolf, Phys. Rev. D, {\bf 2}, 2161 (1970). 
\bibitem{AgaWol70c} G. S. Agarwal, E. Wolf, Phys. Rev. D, {\bf 2}, 2187 (1970). 


\end{thebibliography}
\end{document}